# Optimizing Flux Method Growth of Rutile GeO$_2$ Crystals

Avery-Ryan Ansbro [1] and John T. Heron [1,2a)]

[1] University of Michigan, Department of Materials Science and Engineering, 2300 Hayward St., Ann Arbor, MI 48109
[2] The Ferroelectronics Laboratory, University of Michigan, Ann Arbor, MI 48109, USA

[a)] Electronic mail: jtheron@umich.edu

Rutile germanium oxide (r-GeO$_2$) has shown potential for ultrawide bandgap semiconductor applications such as power conversion and UV optoelectronics. Homoepitaxial substrates will be key for achieving phase pure and doped r-GeO$_2$ thin films as synthesis is inhibited by strain associated with substrate lattice mismatch. Initial reports of single crystal r-GeO$_2$ synthesis from a MoO$_3$-Li$_2$CO$_3$ flux have shown mm scale crystals with dominantly (110) faceting. However, fundamental understanding of the synthesis parameters and the ability to tune size and facet are needed. Here, we report on both seeded and unseeded growth of single crystal r-GeO$_2$ across a range of MoO$_3$-Li$_2$CO$_3$ flux compositions. Small variations in Mo concentration can be used to control crystal habit, faceting, and growth rate through variation in precursor complexion, solution viscosity, and GeO$_2$ solubility. While seed size and seeded growth rates are optimized in 40% Mo solutions, aspect ratios and seeded growth volumes are maximized in 41.5% Mo solutions without sacrificing faceting. Increased Mo concentration leads to polycrystallinity and isotropic growth. These results enable faster and tailored growth of r-GeO$_2$ crystals using the flux growth method.

# I. INTRODUCTION

In the effort to increase the efficiency of power electronics, the need for identifying new dopable, high thermal conductivity, ultrawide bandgap (UWBG) materials to replace traditional group IV semiconductors has emerged. A bandgap higher than that of 3.4 eV in materials such as GaN are typically paired with low dielectric constants, enhanced breakdown field strengths, and higher frequency operation ranges than typical devices which contribute to decreased power loss[1, 2]. Challengingly, with increased bandgap comes higher dopant ionization energies and often asymmetric dopability (unipolar dopability). The resultant dopant asymmetry limits the range of devices that can be produced from these materials[2-4]. Additionally, alternative candidate materials, such as $Ga_2O_3$, are poor thermal conductors due to monoclinic symmetry, a key detriment to operational power electronics.

Subsequently, interest in rutile $GeO_2$ has emerged in recent years. With a predicted direct band gap of around 4.68 eV, r-$GeO_2$ possesses both a relatively high dielectric constant between 12.2 and 14.5 and, theoretically, ambipolar doping despite its ultrawide bandgap[1,3]. The rutile structure is the most stable of several polymorphs at conventional temperatures and additionally is structurally dense, resistant to various acids, and a relatively good thermal conductor[1].

While r-$GeO_2$ is stable at room temperature, during synthesis it is in energetic competition with quartz and glass polymorphs. For thin films in particular, this challenge is enhanced, where strain conditions resultant from lacking substrates with adequate lattice or structural matching can further destabilize the rutile phase[4,5]. A common strategy to combat this challenge for thin films has been through controlling the epitaxial strain by careful substrate selection, a buffer layer, or Sn alloying. Both $Al_2O_3$ and $TiO_2$ substrates have been employed to grow r-$GeO_2$, though typically results in polycrystalline, amorphous, or defect rich films (from relaxation) due the large epitaxial strain[6]. Alloyed $Sn_xGe_{1-x}O_2$ has shown robust phase stability and the control of rutile orientation as a thin film when grown by physical vapor methods, but deteriorates into a glass or multiphase film when Ge concentration becomes high[7, 8]. This limits our ability to fully characterize the properties and ambipolar dopability of pure phase r-$GeO_2$ films.

A potential solution to this problem has been the production of r-$GeO_2$ single crystal substrates. Niedermeier et. al. was able to synthesize r-$GeO_2$ using a $Li_2CO_3$-$MoO_3$ flux and Chae et. al. utilized seeded growth with this method to produce (110) faceted single crystals reaching around 5 mm in length[2,9]. Despite this, early flux growth studies took weeks to create crystals large enough to be practically usable due to the slow cooling rates from elevated temperatures. Additionally, the synthesized crystals were often dark in color due to oxygen vacancies and Mo contamination, which can influence doping and carrier transport[9]. Galazaka et. al. and Goodrum also produced single crystalline $GeO_2$ using the Czochralski method with top seeded growth, exploiting a eutectic between $Li_2CO_3$ and $GeO_2$[10,11]. In particular, Galazaka et. al. produced crystals that could be up to 15 mm in size and electron doped. Antimony doped crystals in particular showcased relatively high mobilities and carrier densities above $10^{20}$/$cm^3$ [10]. Pragmatically, equipment for this technique is somewhat expensive and not necessarily accessible for many labs. It is worth investigating how to produce single crystals with less expensive methods and understand processes and trade-offs regarding the kinetic and thermodynamic factors that will govern the phase, faceting, and density of included defects.

Towards this end, we show that flux composition can be used to control facet preference, growth rate, and inclusion of flux impurities. Through minor alterations in Mo concentration within the flux, crystal habit can be shifted from highly faceted, long, and acicular to small, round, and relatively unfaceted. Without seeding, relatively flat and prismatic single crystals can be grown consistently with the (110) orientation as large as 3 x 3 x 0.5 mm$^3$ using a 40% Mo flux solution. Beyond this, seeded crystals can be grown as large as 4.3 x 2.7 x 1.5 mm$^3$ after only on seeded growth cycle. Additionally, we determine that seeds grow primarily grow between 1000-800 °C due to early growth nucleation, shortening the required growth duration from 8.8 to less than 4 days. This ability to control crystal morphology and growth rate may enhance the practical accessibility of bulk r-GeO$_2$ for use as substrates and other applications where habit control may be helpful.

## II. EXPERIMENTAL

### A. *Flux Growth Method*

Li$_2$CO$_3$ (Fisher Chemical, ≥99%), MoO$_3$ (Thermo Scientific, 99.95%), and hexagonal GeO$_2$ (Strem Chemicals, 99.999% Ge) were used as precursors for crystal growth. Detailed composition of each flux composition can be found in Table 1. The concentration of GeO$_2$ was maintained at 4% of the predicted mass of the flux after Li$_2$CO$_3$ decomposition in all flux variations. Powders were loosely placed in a covered, 100 ml platinum crucible for heating. For seeded growths, three single crystal GeO$_2$ samples were added to the powder flux. Only 40-43% Mo flux solutions were examined. Seeds were chosen to have as similar length and scales and volumes as possible to examine if seed origin affected the quality of growth.

In all experiments, samples were heated from room temperature to their maximum temperature of 1000 °C at 100 °C/hr, where they were maintained for approximately 1-2 hours to ensure dissolution of the $GeO_2$ powder. Maximum heating temperature was chosen as 1000 °C rather than 980 °C as seen in previous work, as this small increase in temperature doubled initial seed sizes (Figure S1) [2, 9]. Crucibles were then cooled to 600 °C at 2 °C/hr, then quenched to room temperature. In attempts to understand supersaturation and gauge rate of seeded growth, some growths were cooled only to 800 or 700 °C instead. Crystals were retrieved and cleaned manually using warm water after crucibles cooled to room temperature. When possible, the volume and dimensions of crystals was obtained with the support of scales and, sometimes, optical microscopy.

A 45% Mo solution without $GeO_2$ inclusion held at 1000 °C for three hours was used to clean the crucible between growths. Crucibles were only replaced when growths degraded, which generally involved either a noticeably reduced size of seeded crystals or a drastic decrease in collected crystals. In these cases, growths were repeated where possible to determine actual behavior of the system.

TABLE I. Composition of varying flux conditions used for growth of r-$GeO_2$ crystals.

|  | 37% Mo | 40% Mo | 43% Mo | 45% Mo | 47% Mo | 50% Mo | 53% Mo |
|---|---|---|---|---|---|---|---|
| $Li_2CO_3$ | 6.557 g | 5.991 g | 5.523 g | 5.144 g | 4.837 g | 4.402 g | 3.994 g |
| $MoO_3$ | 15.002 g | 15.561 g | 16.149 g | 16.398 g | 16.710 g | 17.150 g | 17.546 g |
| $GeO_2$ | 0.706 g | 0.72 g | 0.735 g | 0.739 g | 0.746 g | 0.757 g | 0.766 g |

### B. *Characterization*

X-ray diffraction (XRD) was primarily used to determine crystallinity and faceting of growths with the Rigaku Smartlab diffractometer having a Cu Kα wavelength of 1.541 Å. Bulk measurements were performed with the Bragg-Brentano focus geometry to determine average level of faceting and crystallinity of all crystals. When possible, larger and more highly faceted crystals were examined with a parallel beam geometry and a Ge monochromator to clarify their single crystal nature and growth direction. Preferences for each growth were determined using comparison with polycrystalline diffraction pattern references found on materials project and the crystallography open database. Optical and scanning electron microscopy (SEM) were used to determine the size and shape of grown crystals more accurately. Average crystal shape for each growth condition can be viewed in Supplementary Figure 1. Mass was only recorded for crystals produced in seeded growth experiments.

## III. RESULTS AND DISCUSSION

### A. *Unseeded Growth*

Here we make use of the $Li_2CO_3$-$MoO_3$ flux to grow single crystals and evaluate the impact of the flux composition on them. $Li_2CO_3$-$MoO_3$ is a complex forming flux, meaning that $MoO_3$ forms a complex with precursor molecules to initiate solvation[12]. While $MoO_3$ is not structurally analogous to either $GeO_2$ polymorph, the metal cations in $MoO_3$ and r-$GeO_2$ both are octahedrally coordinated with oxygen ions, potentially allowing for an edge sharing complex to be formed between the two compounds. This is proposed for growths with the structural analog $TiO_2$[13]. As the cations in aq-$GeO_2$ (quartz phase) are instead tetrahedrally coordinated, the octahedrally coordinated Mo complex may favor the formation of the rutile structure from this solution. It follows that experimentally, efforts to grow r-$GeO_2$ in $MoO_3$ systems have rarely reported quartz contamination unless sufficiently high temperatures are reached and efforts to grow aq-$GeO_2$ in complex forming fluxes (with octahedrally coordinated constituents) have often resulted in r-$GeO_2$ contamination[2-3, 14]. While $Li_2CO_3$ is typically added to reduce solution viscosity and volatility, both of which can reduce the quality of crystal growth if too high[15]. It initially decomposes into $Li_2O$, releasing $CO_3$ into the solution. This is ideal when seeking to grow r-$GeO_2$, as the production of $CO_3$ has been shown to disrupt the formation of glass bonds, reducing the stability of the amorphous polymorph[10]. Considering that any carbonate would also possess this property, this would suggest that any $X_2CO_3$-$MoO_3$ (X = period I element) containing flux would be suitable for growing r-$GeO_2$. Recent work from Galazka et. al. shows that $Li_2CO_3$ was superior to other alkali carbonates in growing large crystals on its own and still provides a reasonably low melting temperature when combined with $MoO_3$ suitable for crystal growth near a eutectic, therefore it is reasonable to maintain in our flux[10,16].

Generally, the ratio between flux components is chosen close to the eutectic, minimizing the melting point of the solution. Lower temperature growths are less energy intensive. While $GeO_2$ is not insoluble in $Li_2CO_3$, $GeO_2$ complexion (and decomplexation) with $MoO_3$ is the main mechanism of both solvation and crystal growth in this system[13]. It would follow that varying its composition within the flux should alter growth and solvation dynamics and complex mobility for tuning the crystal growth.

Thus, we begin our study by evaluating the efficacy of r-GeO$_2$ synthesis under varying flux compositions. Mo concentration varied between 37-53% to include both hypo- and hyper-eutectic compositions associated with the Li$_2$CO$_3$-MoO$_3$ system. (Figure 1a, Table 1). It is expected that the "real" phase diagram, one that is inclusive of the GeO$_2$ contribution, does not match the Li$_2$CO$_3$-MoO$_3$. Nonetheless, it can still provide a useful framework for understanding the system, predicting melting range of the flux, and designing the experiment. Total mass of flux, GeO$_2$ concentrations, temperature ramp, and environmental conditions for growth were maintained between each experiment for consistency. Further details of the unseeded flux growth can be found in Section I.

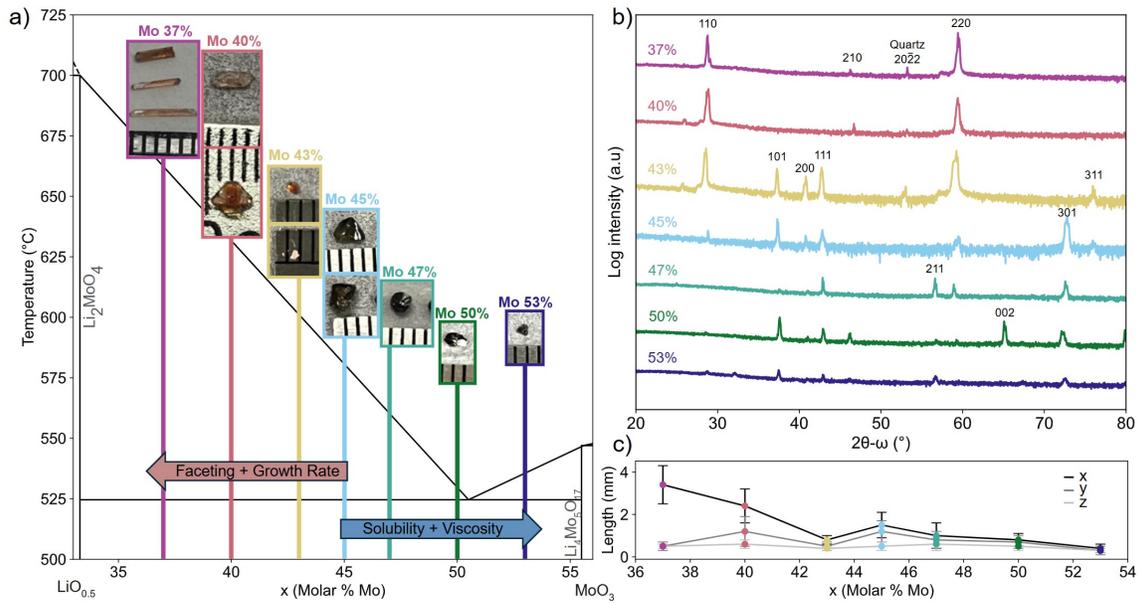

FIG. 1. Unseeded crystal morphology and faceting with flux composition. a) Shows the resulting crystal habit and size from the flux growth at the corresponding flux composition. Phase diagram is redrawn from Moser et. al[16]. All growths were otherwise performed at the same conditions. All ruler notches in the optical images are in mm

interval. Increasing the Mo concentration in the flux increases the Ge solubility and flux viscosity while decreasing the crystal faceting and growth rate. b) XRD of all recovered crystals grown at each flux condition. The crystals become dominantly (110) faceted as the Mo concentration is reduced. c) Averaged aspect ratios of the crystals as a function of flux composition. At low Mo concentration, the crystals are more anisotropic.

It was found that even minor changes in flux composition could alter the habit, growth rate, and contamination rates of the growths. Figure 1a presents optical images of the resultant crystals at the corresponding flux composition, superimposed with the $LiO_{0.5}$-$MoO_3$ phase diagram. With a reduction in Mo concentration, it was found that average crystal size increased, with crystals reaching up to 5 mm and, on average 4 mm in length when grown in solutions containing 37% Mo. Additionally, faceting greatly increased, with 37 and 40% Mo growths containing primarily crystals of the (110)-orientation (Figure 1b). Faceting was consistently large enough that individual crystals could be affirmed as single crystals (Figures S2, S3). As a result, crystals in the 40% Mo growths took on a flat and quadrilateral habit while 37% Mo growths were acicular and frail (Figures 1 and S2). When Mo concentrations were instead increased, growths became increasingly spherical as growth direction became less preferred, evidenced by the range of growth planes observed in XRD spectra of all collected crystals (Figure 1b-c). Individual crystals also consistently reduced in size and darkened in color, the latter suggesting an increased Mo concentration.

These results could potentially be explained with changes in solution supersaturation. Nucleation of new crystallites is more likely to occur under higher levels of supersaturation while growth of larger crystals is supported by lower levels of supersaturation[12,17,18]. Assuming sufficient solution diffusivity and cooling rate, a decreased Mo concentration could result in a lowered supersaturation, hence larger crystals. A shift from acicular to plate like to variably shaped geometries can also be observed as supersaturation is increased within other systems, as an increased precipitation rates is often paired with anisotropic growth[19]. Challengingly, for systems cooled at the same rate with the same percentage of solute dissolved per flux volume, the level of solubility would initially dictate the level of supersaturation during a given set of conditions. That said, since it is expected that $GeO_2$ is more soluble in $MoO_3$ than $Li_2CO_3$, there is an argument for a reduction in supersaturation with increasing Mo concentration instead.

Variation in flux viscosity resulting from the high viscosity of $(l)MoO_3$ may better explain the observed behavior[20]. As solution viscosity increases, $GeO_2$ and complexed materials' ability to travel throughout the solution is minimized. Growth becomes kinetically rather than thermodynamically limited, resulting in an increase in random nucleation and a reduction in relative stability difference between individual growth orientations. This results in round, prismatic crystals with a low degree of faceting and minimizes capacity for growth at extremes. Decreasing viscosity enhances mobility, shifting growth to the most stable crystal orientations. With lower viscosity solutions, as seen with 37% Mo, growths become highly directional. Due to lack of growth in other directions, however, these crystals become frail and more viable for microscale wires than as substrates. These results are consistent with what has been observed in other complex forming fluxes[13, 21]. Still, considering that variations in flux chemistries are small, viscosity being the only important factor in habit changes and growth rate needs further investigation.

Considering its yield of numerous, relatively large seeds (≥ 2 mm), crystal habit, and tendency for faceting in the (110) direction, 40% Mo is the most ideal condition for unseeded growth when considering desire for substrate production. Still, crystal sizes are still too small to be practical on their own. To achieve larger crystals with controlled facet orientation, we next explore the influence of seeded crystal synthesis.

## B. Seeded Growth

### 1. $GeO_2$ Solubility

To further understand how solubility affects growth rate and habit as well as determine the quantity of time required for seeded growths, the quenching temperature was varied between 800-600 °C. Seeds were chosen to be as similar as possible to one another for these growths. For consistency, only seeds grown using a 40% Mo flux were chosen for higher accuracy in comparison. Only Mo concentrations between 40-43% were examined, since seeds grown in these regimes minimized Mo contamination, had low apparent variation in crystal orientation, and produced the most ideally shaped seeds.

Measuring precipitated crystals after quenching at various temperatures can provide a crude estimation of $GeO_2$ solubility in solution given our specific growth conditions (Figure S2). While this isn't representative of true solution capacity, it provides a means to compare supersaturation between solutions and understand variation in growth rate for both unseeded and seeded growth. Unexpectedly, solubility does not vary significantly between Mo 40-43% growths despite crystal habit and faceting varying notably between systems. Notably, we see large variations in yield from the 41.5% system, though we expect that this could be related to the degradation of platinum crucibles over the course of repeated experiments. Based on investigation of rutile $TiO_2$ solubility in the $Li_2O$-$MoO_3$ flux, we still expect solubility to increase with increasing $MoO_3$ concentration despite these results. Alongside our results, if solubility trends follow $TiO_2$, the challenge differentiating solubility between different conditions here indicates that true variation is likely small between different solutions[13]. This suggests the variation in habit seen during both seeded and unseeded growth and the tendency towards polycrystallinity during seeded growth with increased Mo concentration is not likely related to levels of supersaturation.

2. *Optimizing Directional Lateral vs Volumetric Growth*

We next worked to determine the most optimal conditions for seeded growth. We once again chose to focus on only 40-43% Mo solutions (will be referred to as "seeded growth solution"). We also varied seed choice (which will be referred to as "seed origin") between those produced in the 40%-43% Mo methods to see if seed habit impacted growth rate or quality to a significant degree. Level of crystallinity and faceting direction could not always be determined for individual seeds considering crystals were often too small and, particularly for those chosen from 43% Mo solutions, unfaceted for successful analysis. Seeds from the Mo 40% method was assumed to be dominantly faceted across the (110) plane based on bulk XRD measurements, seeds for the 43% Mo solution were expected to vary in faceting. For consistency, seeds were chosen to have similar starting dimensions and/or volumes so that growth rates could be effectively compared. Since individual seeds could often not be traced after growth and were chosen to be as similar as possible, reported seed origin dimensions are averaged. Only the largest grown crystal's dimensions are reported after seeded growth to determine maximum growth capacity of selected flux chemistries, though the two largest crystals in each considered growth are reported for consistency in Table S1.

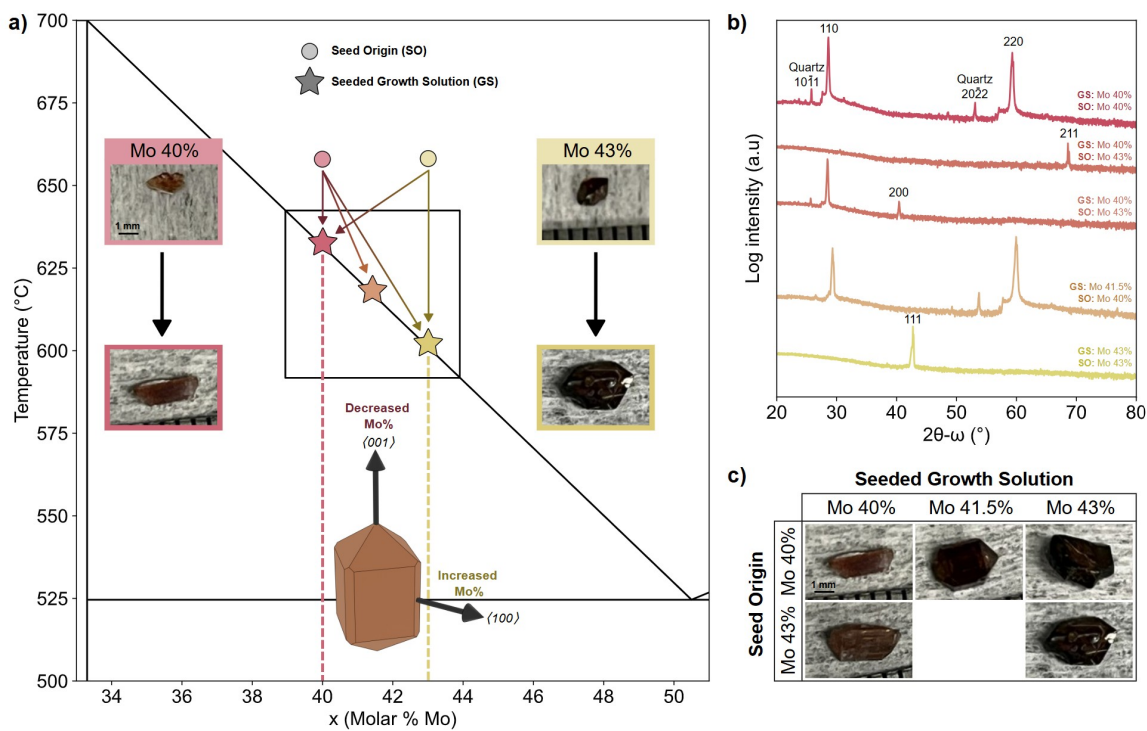

FIG. 2. Seeded Growth. a) Seeds grown using Mo 40% and 43% compositional fluxes were further grown in Mo 40-43% flux solutions to determine if seed type and/or growth condition effected resultant crystal size and facet. The Wulff is inset in this figure. Paired with b) which shows XRD associated with largest grown (and otherwise relevant) crystal(s) from each solution where possible and c) which shows largest grown crystal under each condition, it is observed that (110) faceting increases from seeded growth when a flux with a lower Mo concentration is used. The resultant crystal volume tends to be larger for higher Mo concentration by promotion of multidirectional growth. Unlabeled, small peaks in b) belong to the $Li_2MoO_4$. Spectra from seeds originating from 40% Mo conditions grown in the 43% Mo solution is not represented due as abnormal crystal shapes do not allow for measurement. See Tables 1 and S1 for quantitative descriptions.

TABLE 2. Results of seeded growth under varying conditions. Average dimensions of starting seeds and maximum dimensions of grown seeds are reported. Initial mass of seed material was not recorded.

| Growth Condition (% Mo) | Seeds (% Mo) | Ave. Starting Size (mm³) | Final size (mm³) | Total growth (mm³) | Faceting | Mass (mg) |
|---|---|---|---|---|---|---|
| 40% | 40% | 2.6 x 1.1 x 0.8 | 4.0 x 1.5 x 1.0 | 1.4 x 0.4 x 0.2 | (110) and quartz | 23.3 |
| | 43% | 2.0 x 1.5 x 1.0 | 4.0 x 2.0 x 1.2 | 2.0 x 0.5 x 0.2 | (110), (200), (211), quartz | 49.5 |
| 41.5% | 40% | 3.0 x 1.0 x 1.0 | 4.3 x 2.7 x 1.7 | 1.3 x 1.7 x 0.7 | (110) and quartz | 94.1 |
| 43% | 40% | 2.3 x1.6 x 0.9 | 4.0 x 3.1 x 2.6 | 0.7 x 1.5 x 1.7 | n/a | 92.1 |
| | 43% | 2.1 x 1.2 x 1.1 | 4.0 x 2.3 x 1.0 | 0.9 x 1.1 x 0.1 | (111) and n/a | 55.6 |

As described and visualized in Table 2 and Figure 2, similar trends in contamination and faceting were observed in seeded growths. Challengingly, we observe that most grown crystals are polycrystalline. Concerningly, quartz contamination is more likely to be observed in solutions with lower Mo concentrations. This was also observed in unseeded growth, particularly when temperature range for growth was shifted from 980 to 1000 °C (Figure S1). We expect temperature to play a large role in the development of multiphase seeds and expect that this could be eliminated or made less likely by lowering maximum growth temperatures to 980 °C for seeded growth.

It is also noted that seeded growth in a 40% Mo solution often produced more lightly colored crystals dominantly faceted in the [110] direction while seeded growths in 43% Mo solutions were darker in hue and inconsistently faceted. Additionally, seeded crystals grown in a 43% Mo solution often became polycrystalline or irregular in shape regardless of seed origin. When faceting of grown crystals in 43% Mo solutions could be measured, it was often of the (111) plane, though previous work suggests that single crystal, (110) faceted crystals can be grown using this chemistry[2,3]. While lower Mo concentrations continued to promote growth along the c-direction during seeded growth, it did not guarantee that the resultant crystal would be 110 faceted. When seeds originating in Mo 43% solution were further grown using a 40% Mo solution, the two largest crystals were dominantly faceted in the (110) and (221) directions respectively (Figure 2b). The (221) faceted crystal also maintained irregular shaping, so it would require more processing if it were to be used as a substrate.

Looking at total growth by direction and volume from initial crystals to final product (Table 2), we observe that rather than faceting becoming somewhat random, growth in the equivalent a and b directions becomes favored as Mo concentrations increases in the flux solution. While viscosity may still contribute to habit variation, these results are strongly suggestive of some interaction between the flux solution and the growth planes themselves. Considering $MoO_3$ is a complex former with $GeO_2$ and is retained in crystals as an impurity after growth, it is strongly suspected that $MoO_3$ selectively adheres to growing planes. Depending on the nature of the interaction, $MoO_3$ adhesion could either increase or decrease the rate of growth of a given plane either through decreasing active sites available for $GeO_2$ attachment or by acting as a transport molecule for enhancing rate of $GeO_2$ attachment to the growing plane. Considering that the (111) planes possess more Ge and O active sites than the (110) plane (and other planes more generally), it is likely that $MoO_3$ attaches to the (111) plane (as well as other more reactive planes), slowing growth, thereby promoting larger facets of otherwise less stable growth planes[15,19,22]. We expect this interaction also contributes to increased Mo concentration in crystals beyond what would be expected just from the increased presence of MoO$_3$ in the solution and solute trapping alone.

While a decreased Mo concentration results in a stronger tendency to maintain faceting along the [110] direction, challengingly, it does not result in larger overall crystals. Comparing mass of grown crystals, a higher Mo solution concentration generally increases the volume of crystals extracted from seeded growth. This is somewhat unexpected, considering that unseeded crystals are generally larger when extracted from a 40% Mo solution. It is possible that, despite MoO$_3$ adhering to active sites, the development of more reactive faceting directions provides more surface area and active sites for growth to occur overall. Reduced presence of MoO$_3$ likely also reduces energy required for precipitation to occur due to less complexion, so it is also possible that lower % Mo solutions suffer from more spontaneous nucleation. Considering the restriction associated with our investigation, it is unclear how total crystal yield varies with solution concentration due to the number of crystals produced in all experiments and immeasurable, though impactful contribution of crucible degradation to spontaneous nucleation rates.

To determine if the % Mo concentration "limit" for consistent production of dominantly (110) faceted crystals or if there was a higher concentration that could achieve better volumetric growth while maintaining sufficient growth isotropy, a flux solution containing 41.5% Mo was examined. Excitingly, grown crystals consistently had a higher volumetric growth rate and produced primarily crystals with dominant (110) faceting. Quartz contamination was still observed here. Resultant from comparatively isotropic growths between 40% and 41.5% Mo solutions, extracted crystals tend to have a smaller aspect ratio that is more typical for substrate usage. This appears to be a more ideal condition for seeded growth than 40% Mo solutions when considering crystal size alone.

TABLE 3. Results of seeded growth after precipitation at 800 °C. Average dimensions of starting seeds and maximum dimensions of grown seeds are reported. Initial mass of seed material was not recorded.

| Mo % | 40 | 41.5 | 43 |
|---|---|---|---|
| $GeO_2$ Precipitated | 60.1% | 57.4 (±22.9) % | 62.2% |
| Starting Seed Dimensions ($mm^2$) | 2.2 x 1.1 x 0.9 | 1.9 x 1.5 x 0.9 | 2.1 x 1.0 x 1.0 |
| Max Dimension ($mm^2$) | 4.0 x 2.0 x 1.9 | 3.0 x 2.3 x 1.3 | 4.1 x 2.8 x 2.0 |
| Average mass (mg) | 26.0 (±11.7) | 22.5 (± 9.6) | 55.6 ± (11.1) |

Finally, we tracked rate of crystal growth as it varied with change in temperature solution chemistry to fine tune growth conditions. As described in Table 3, just over half of $GeO_2$ has precipitated by 800 °C. Notably, crystals in two of four systems are laterally equivalent in size and two are volumetrically. We suspect that the decreased mass and lateral growth seen in the 41.5% solution could be related to degradation of the platinum crucible considering the ordering of experiments.

What is notable about these results is that, by both mass and volume, 40% Mo solutions are statistically identical in both size and volume by 800 °C to crystals allowed to grow until 600 °C. While mass is lower than the maximum expected from a crystal grown in the 43% Mo solution, lateral dimensions are similarly equivalent to what would be expected to extract from a longer growth. Though higher supersaturation levels seen under higher temperature conditions typically favors nucleation over growth, seed crystals act as substrates that greatly reduce energies for nucleation on their surfaces during the first half of seeded growth. It is not surprising that rates of seed growth in most substantial when precipitation rates are at their highest. Spontaneous nucleation is still common in these systems, leading to many crystals being generated by 800 °C despite the presence of seeds. This quantity of excess crystals could only be reduced if a new crucible were used for each growth but not prevented. This means that as the system progresses to lower temperatures where growth becomes favored over nucleation, the quantity of "seeds" available for further growth is numerous. This reduces preference for growth on the original, larger seeds resulting in negligible volumetric growth beyond 800 °C. While this decreases the efficiency of growth overall, this also means that temperatures for these growths are most ideally from 980-800°C, shortening seeded growths from 8.8 days to just under 4 days.

## IV. SUMMARY AND CONCLUSIONS

By adjusting the composition of the $MoO_3$-$Li_2CO_3$ flux, crystal growth habit, crystallinity, contamination levels, and growth rate can be fine-tuned and controlled. Lower Mo concentrations reduce the prevalence of polycrystallinity due to promoting directional growth in the (001) direction. We suspect that this is directly related to $MoO_3$ adhesion to reactive sites on growing crystal planes. Increased Mo concentration promotes growth in equivalent a and b directions, resulting in multidimensional growth within the flux concentrations surveyed. Thus, solution ratio between $Li_2CO_3$ and $MoO_3$ can be used to control crystal habit, faceting and growth rate under both seeded and unseeded growth conditions.

This investigation allows for the optimization of growth conditions for r-GeO$_2$ crystals. 40% Mo solutions seem to be most ideal for producing initial (110) faceted crystals, optimizing large growth rates with a prismatic and plate-like habit resembling substrate dimensions. For producing larger crystals, seeded growth using a 40-41.5% Mo solution and seeds originating from 40% Mo solutions provides conditions that maintain growth sufficient anisotropy to promoting the maintenance of (110) faceted crystals. To avoid quartz contamination and reduce time required for growth, optimal growth conditions are suggested to be between 980-800 °C.

While interactions between flux and growth planes seems to explain our results well, experimental resources prevent adequate investigation into the influence of solubility on growth. It is noted that crucibles contribute greatly to spontaneous nucleation for these growths and that further investigation into cheaper and more durable crucibles would greatly benefit this field. Considering recent advances in GeO$_2$ production, it is unlikely such investigations would be valuable for this system, though may benefit experimentation with other material types[10]. Through this methodology, it is possible to consistently produce sufficient, comparatively inexpensive, substrates in around two weeks (excluding the polishing process). This may make GeO$_2$ substrates more accessible, which could allow research progress investigating the properties of GeO$_2$ and its alloys to proceed before these substrates can be regularly produced through industrial means.

# SUPPLEMENTARY MATERIAL

Review supplementary material for a more rigorous examination of crystal habit, crystallinity, and for size and shape distribution of the largest crystals grown from seeded growths.

## ACKNOWLEDGMENTS

We are grateful for the support of Van Vlack laboratories and the Michigan Center for Materials Characterization for the tools and support for this work. We appreciate and acknowledge feedback from Dr. Wenhao Sun, Dr. Na Hyun Joe, and Ashiq Shawan for their helpful feedback on initial experimental plans. This material is based upon work supported by the National Science Foundation under Grant No. 2328701 and is supported in part by funds from federal agency and industry partners, as specified in the Future of Semiconductors (FuSe) program.

## AUTHOR DECLARATIONS

**Conflicts of Interest**

The authors have no conflicts to disclose.

## DATA AVAILABILITY

Data available on request from the authors and within the article and supplementary materials.

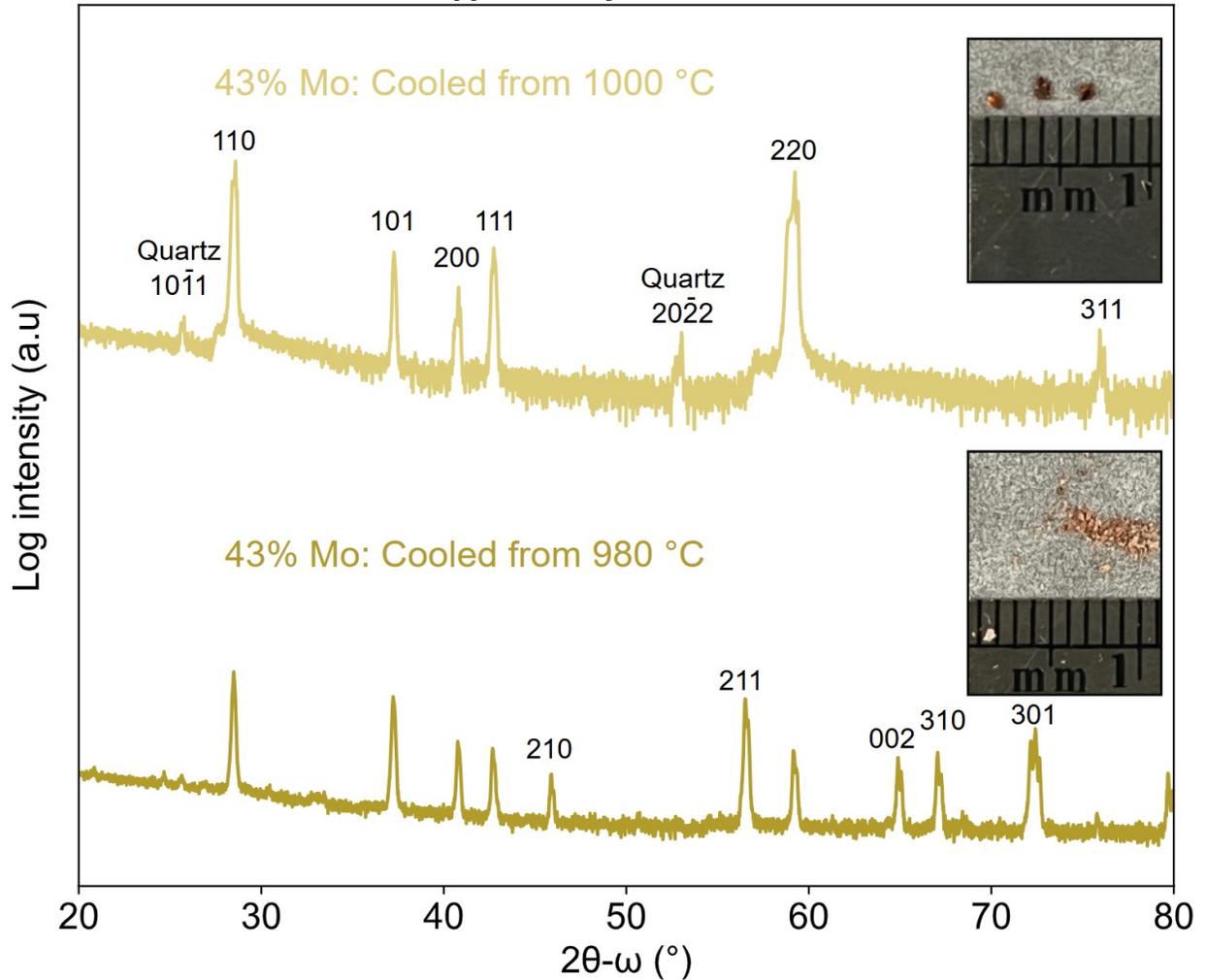

**Figure S1:** Comparison in unseeded growths cooled from 980 °C vs 1000 °C. Crystals cooled from 980 °C are generally ≤ 1 x 1 mm² and present with flake morphology. While small quantities of quartz can be found when cooling from 1000 °C, increased intensity of (110) peaks suggest that grown crystals are highly faceted. Collected crystals are volumetrically larger and prismatic in morphology. This shift may be attributed to the increased rate of solubility change with temperature found at higher temperatures and subsequent variations in supersaturation.

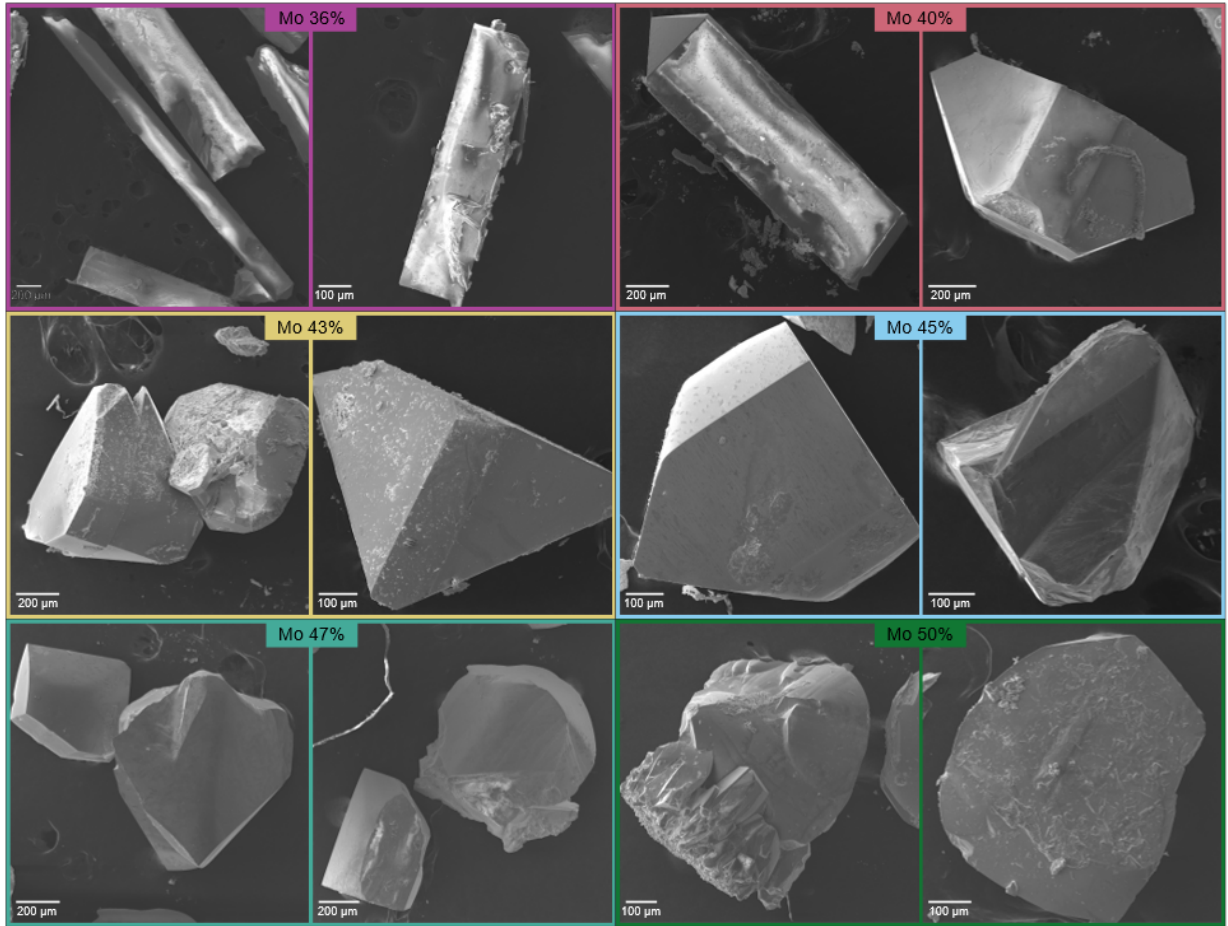

**Figure S2: Variation in seed morphology with variation in flux chemistry.** SEM images of grown seeds showcasing variation of crystal habit under varied flux compositions. Primarily smaller (< 2 mm) were chosen for observation. Small, dendritic crystals atop seeds are solidified flux that was not removed before imaging. Triangular (111) planes can begin to be visualized by growths in 40% Mo, however they do not begin to dominate seeds until growths in 43% Mo. Beyond this, large and flat crystal planes can still be obtained, though XRD is suggestive that these planes are not frequently of the (110) direction. Crystals grown in 50% Mo are visibly polycrystalline.

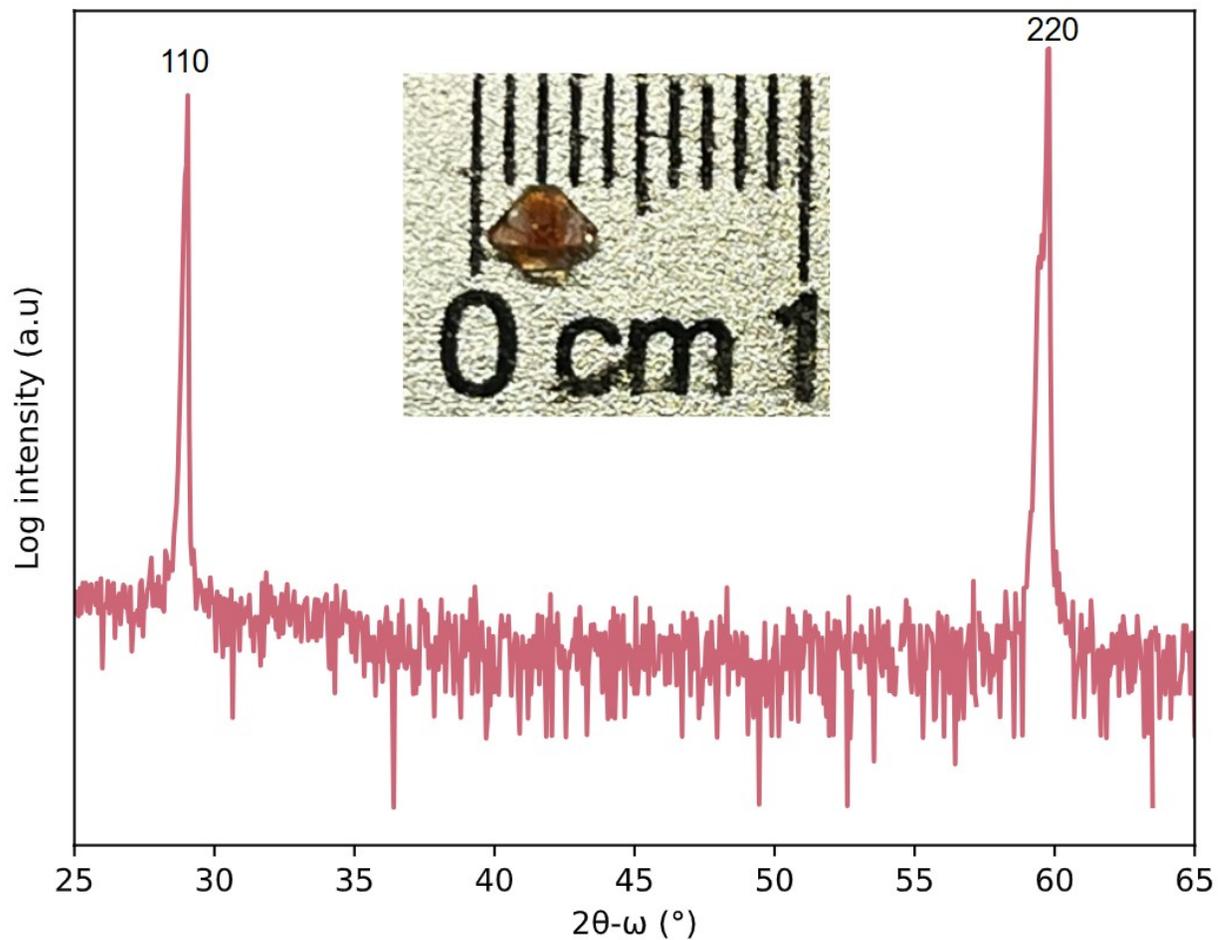

**Figure S3: Crystallinity of Seed Grown in 40% Mo flux solution.** XRD diffraction suggests that the large flat seed is primarily oriented in the (110) direction.

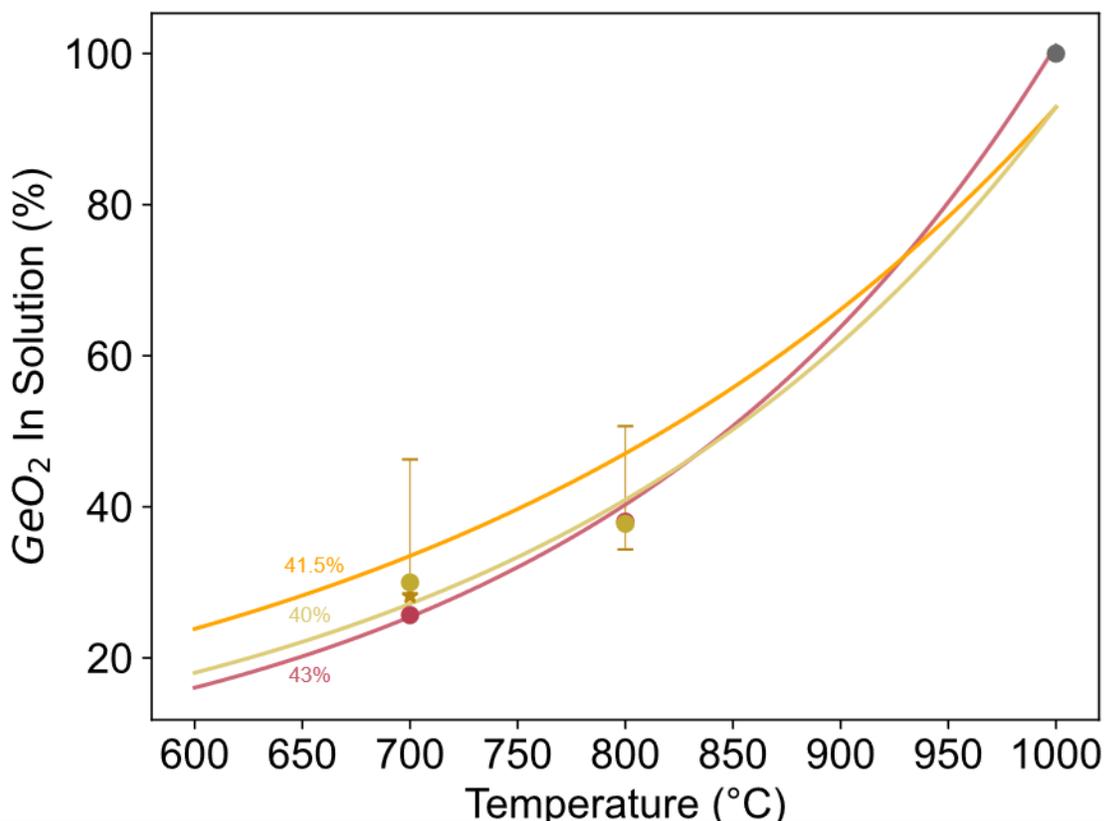

**Figure S4:** Optimization of growth rate though solubility estimation. Through crystal collection at the end of growth, we estimated amount of $GeO_2$ remaining in solution after cooling to varying temperatures. When fit to an exponential curve, this can be used to approximate solubility and compare supersaturation levels between solutions. Challengingly, it does not appear as if there is any statistical difference between solubilities for Mo concentrations between Mo 40-43%.

**Table S1**

| Growth Solution-Seed Origin | Dimensions (mm³) | Morphology | Faceting |
|---|---|---|---|
| Mo 40-40% | 4.0 x 1.5 x 1.0 | Prismatic, flat | 110 and quartz |
| | n/a | n/a | n/a |
| Mo 40-43% | 4.0 x 2.0 x 1.2 | Prismatic, flat | 110, 200, quartz |
| | 3.0 x 2.5 x 2.0 | Irregular, rounded | 211 |
| Mo 41.5-40% | 4.3 x 2.7 x 1.7 | Prismatic, flat | 110, quartz |
| | 4.7 x 2.4 x XX | Prismatic, flat | 110, quartz |
| Mo 43-40% | 4.0 x 3.1 x 2.6 | Irregular, flat/twined | n/a |
| | 3.8 x 2.1 x 1.0 | Irregular, rounded | n/a |
| Mo 43-43% | 4.0 x 2.3 x 1.0 | Prismatic, rounded | 111 |
| | 3.0 x 2.5 x 1.0 | Irregular, rounded | n/a |